# The Dielectric Response of $La_{0.5}Ca_{0.5-x}Sr_xMnO_3$ ($0.1 \leq x \leq 0.4$) Manganites with Different Magnetic Ground States

Indu Dhiman, S. K. Deshpande[1], and A. Das[*]

Solid State Physics Division, Bhabha Atomic Research Centre, Mumbai – 400085, India.

[2]Inter University Consortium for DAE Facilities, Mumbai Centre, Mumbai – 400085, India.

## Abstract

The dielectric behavior of half doped manganites $La_{0.5}Ca_{0.5-x}Sr_xMnO_3$ ($0.1 \leq x \leq 0.4$) with varying magnetic ground states has been studied. The real part of relative permittivity as a function of temperature $\varepsilon'(T)$, exhibits a maximum around the ferromagnetic ($T_C$) and charge ordering transition ($T_{CO}$) temperatures accompanied with high dielectric losses. The activation energies obtained for x = 0.1 and 0.3 samples below $T_{CO}$ are the same ~ 0.12eV, whereas the relaxation time constant varies in the range $2.8 \times 10^{-9}$ s – $6.03 \times 10^{-11}$ s. In contrast to samples having $x \leq 0.3$, for x = 0.4 doping the dielectric permittivity exhibits a strong temperature dependence in the vicinity of magnetic phase transitions. This behavior may be correlated with the presence of competing magnetic interactions (magnetic polarons) close to the magnetic transitions.

PACS: 75.47.Lx, 77.22.Ch, 75.85.+t

Keywords: Manganites, Dielectric properties, Mulitferroic

# INTRODUCTION

The study of multiferroic materials with coexisting magnetic and ferroelectric ordering is attracting considerable interest.[1,2] The multiferroic nature of manganites has been reported recently in several systems such as, $RMnO_3$ oxides (R = Y, Gd, Tb, and Dy)[3-6], $Gd_{0.5}Sr_{0.5}MnO_3$[7,8], $R_{0.5}Ca_{0.5}MnO_3$ (R = La, Tb, Dy, Er, and Nd)[9-12], $Pr_{0.6}Ca_{0.4}MnO_3$[13,14], and $Pr_{0.7}Ca_{0.3}MnO_3$[14], $La_{0.67}Ca_{0.33}MnO_3$.[15] The half doped charge ordered manganites, in particular exhibit a large electronic polarization.[4-6] The magnetic field is observed to have a strong influence on the dielectric properties, associated with the phase separation behavior evident in charge ordered manganites.[14,15] In addition, the electric field gradient experiments suggest the existence of ferroelectric domains in charge ordered regime.[16] The coexistence of ferroelectricity and magnetism in charge ordered manganites is explained to be a result of canted antiferromagnetic state which drives the site centered charge ordering behavior to bond centered charge ordering behavior.[10] The nature of charge ordering behavior in $La_{0.5}Ca_{0.5}MnO_3$ compound is debated between site centered and bond centered. The previously reported dielectric study on this compound shows an anomalous behavior close to the charge ordered transition temperature.[9] In the present study, we have investigated the dielectric behavior of $La_{0.5}Ca_{0.5-x}Sr_xMnO_3$ (x = 0.1, 0.3, and 0.4) series, exhibiting varying magnetic ground states. Previously reported neutron diffraction studies on these compounds show the CE – type antiferromagnetic structure is stable for x < 0.3.[17,18] At x = 0.3, CE – type ordering coexists with an A – type antiferromagnetic phase. At x = 0.4, A – type antiferromagnetic order replaces the CE – type state. Evidence of long range ferromagnetic ordering in the temperature range of 180 – 250K is observed only in case of x = 0.4 sample. All the compounds having x ≤ 0.4 exhibit an insulating behavior.

**EXPERIMENT**

The polycrystalline samples were synthesized by conventional solid-state reaction route and the method preparation has been reported elsewhere.[17] Temperature dependent dielectric measurements were carried out using Novocontrol alpha impedance analyzer in the frequency range of 100Hz to 5MHz, with rms ac voltage of 1V. Samples in the form of pellets were sandwiched between the electrodes of Novocontrol BDS1200 sample cell. The sample temperature was varied between 293K and 123K using the Quatro cryosystem with liquid nitrogen for cooling.

**RESULTS AND DISCUSSION**

The structural and magnetic phase diagram for $La_{0.5}Ca_{0.5-x}Sr_xMnO_3$ ($0.1 \leq x \leq 0.5$) series obtained from unpolarized neutron diffraction study has been reported previously.[17] We have observed that with progressive increase in Sr doping the CE – type antiferromagnetic phase is suppressed and ferromagnetic phase is favored. The CE – type antiferromagnetic phase is observed for x = 0.1, 0.2, and 0.3 samples with antiferromagnetic transition temperatures 150, 100, and 75K, respectively. At x = 0.4, the CE – type antiferromagnetic structure is fully suppressed and A – type antiferromagnetic phase is observed with the transition temperature $T_N$ = 200K. Further increase in Sr doping at x = 0.5, the long range ordered ferromagnetic phase is established for all temperatures below 310K. Structurally, the Sr doped samples having x ≤ 0.3 crystallize with orthorhombic structure (space group *Pnma*) and x = 0.4 crystallizes with two orthorhombic phases in the space group *Pnma* and *Fmmm*. In the present study, we have chosen a few representative compounds from $La_{0.5}Ca_{0.5-x}Sr_xMnO_3$ (x = 0.1, 0.3, and 0.4) series having distinct magnetic structures. In figure 1(a) magnetization as a function of temperature M(T), for

x = 0.1, 0.3, and 0.4 samples is displayed.[17] All the three samples exhibit magnetic transitions attributable to the onset of charge ordering ($T_{CO}$), ferromagnetic ($T_C$), and antiferromagnetic ($T_N$) states, similar to x = 0 sample.[19] Figure 1(b) shows the temperature dependence of real part of relative permittivity $\varepsilon'(T)$ for all the three samples at frequency ~ 12kHz. For x = 0.1 and 0.3 samples with CE – type antiferromagnetic state, the $\varepsilon'(T)$ exhibits a maximum at 230 and 280K, respectively. This maximum coincides with the maximum in M(T), evident in figure 1. Similar behavior of $\varepsilon'(T)$ having a maximum close to $T_{CO}$ is observed in x = 0 [9] and other charge ordered manganites.[20-22] The charge ordering transition in these compounds is found to be accompanied by a structural transition from orthorhombic phase having space group *Pnma* to a lower symmetry monoclinic phase with *P2$_1$/m* space group. This space group accounts for the two inequivalent sites for $Mn^{3+}$ and $Mn^{4+}$ ions, which are created as a result of charge ordering leading to higher distorted octahedra. The presence of Jahn Teller lattice distortions in addition to strong electron phonon interactions lead to a canting instability of antiferromagnetic state. This behavior favors formation of spin dimers that drives charge ordering from site centered to bond centered which leads to the appearance of ferroelectric properties in charge ordered manganites.[10] In these samples, using diffuse neutron scattering studies we have found evidence of short range ordering at temperatures above $T_{CO}$.[18] The short range ordering is found to grow on lowering of temperatures and is suppressed below the magnetic ordering temperatures. The shallow maximum in $\varepsilon'(T)$ for T > $T_C$ correlates with this. The x = 0.4 sample exhibits an A – type antiferromagnetic state and ferromagnetic phase in the intermediate temperature range of 180 – 250K. The variation of $\varepsilon'(T)$ exhibits a strong temperature dependence for T > 200K, close to the ferromagnetic transition and its value becomes negative (not shown in the figure) above 250K. This behavior is significantly different as compared to x = 0.1 and 0.3 samples.

Additionally, in antiferromagnetic state the absolute value of $\varepsilon'$ is higher in case of x = 0.4 sample indicating the influence of nature of magnetic ordering on $\varepsilon'$. The dielectric response close to the transition temperature may be correlated with competing ferromagnetic and antiferromagnetic interactions present near the transition temperature. This behavior is in agreement with earlier reported polarized neutron scattering study on these samples, where magnetic polarons of antiferromagnetic and ferromagnetic nature coexist above the transition temperatures.[18] Magnetic field dependent dielectric study on $La_{2/3}Ca_{1/3}MnO_3$ system reveals the influence of phase separated state on the dielectric behavior near the magnetic transitions.[15] The temperature dependence of imaginary part of relative permittivity, $\varepsilon''(T)$, is shown in the inset to figure 1(b). The primary contribution to $\varepsilon''$ is governed by dc conductivity ($\sigma_{dc}$) and is expressed as $\varepsilon'' = \sigma_{dc}/\omega\, \varepsilon_0$, where $\omega$ is the angular frequency and $\varepsilon_0$ is the dielectric permittivity of free space. At frequency $\omega \approx 12kHz$, for Sr doping at x = 0.1 the $\varepsilon''(T)$ exhibit a rise with increasing temperature. This behavior is consistent with earlier reported transport studies, wherein conductivity increases with increasing temperature. For higher Sr substitution (x = 0.3 and 0.4), $\varepsilon''(T)$ shows similar temperature dependence. At a given temperature, as a function of Sr doping, enhancement in $\varepsilon''(x)$ indicate the favoring of metallic behavior, in corroboration with transport study. Figure 2 shows the Argand plane plots (imaginary part of impedance (Z″) versus real part of impedance (Z′)) for $La_{0.5}Ca_{0.5-x}Sr_xMnO_3$ (x = 0.1, 0.3 and 0.4) at 123K. One single semicircle has been observed for all three samples in frequency range of 100Hz – 1MHz. The size of the semi circle is dependent upon the size and number of grains present in the sample, while the diameter of semicircle corresponds to resistance from the grains.[23,24] The presence of single semi circle in the studied frequency range for all three samples corresponds to intra grain contribution to conductivity. As a function of Sr doping, the diameter of the semicircle reduces. The large decrease in diameter with composition indicates the reduction in resistivity from grains and increase in conductivity, which is in agreement with our previously reported dc resistivity studies.[17] In figure 3, $\varepsilon'$ as a function of frequency $\varepsilon'(\nu)$, for x = 0.1 sample at several temperatures is shown. The frequency dependence of $\varepsilon'$ can be described by Debye equation expressed as,[25]

$$\varepsilon' = \varepsilon_\infty + \frac{\varepsilon_s - \varepsilon_\infty}{1 + (i\omega\tau)^{1-\alpha}} \qquad (1)$$

where, $\varepsilon_\infty$ and $\varepsilon_s$ are the respective value of relative permittivity at high and low frequency, $\omega$ is the angular frequency, $\alpha$ is a measure of the distribution of relaxation time and $\tau$ is relaxation time. The thermal dependence of $\tau$ is Arrhenius type, $\tau = \tau_0 \exp(E_A/k_BT)$, where $\tau_0$ is a constant and $E_A$ is the activation energy. The fit at various temperatures for x = 0.1 sample is shown in figure 3. The relaxation time $\tau$ is obtained by fitting $\varepsilon'(\nu)$ to equation (1) and is plotted as a function of 1000/T in figure 4 for x = 0.1 compound. It is evident in figure 4 that $\tau$ obeys Arrhenius law in the temperature regime of 123 – 253K. The fitting parameters obtained are $E_A \approx$ 0.12eV and $\tau_0 \approx 2.8 \times 10^{-9}$s. Similar to x = 0.1 sample, for x = 0.3 sample the frequency dependence of $\varepsilon'$ could be described by equation 1. Figure 5 shows the relaxation time $\tau$ versus 1000/T along with Arrhenius fitting for x = 0.3 sample in the range of 123 – 203K. The obtained fitting parameters are, $E_A \approx 0.121$eV and $\tau_0 \approx 6.03 \times 10^{-11}$s, showing no significant change in comparison with x = 0.1 sample. A similar analysis for x = 0.4 could not be carried out because of the negative values of $\varepsilon'(\nu)$ in several temperature range. Negative capacitances (and therefore dielectric constants) have been observed in similar systems[26] and has been primarily attributed to delocalization of trapped carriers. Additional effects like contacts, interfaces, space charge etc. may also lead to negative capacitances. In the case of x=0.4, the conductivity at low temperatures is highly enhanced as compared to charge ordered x=0.1 and 0.3 compositions. Under the application of an electric field the carriers hop between localized states and cause an inductive effect giving rise to negative capacitances. In a dielectric measurement the net capacitance is a sum of effect from the bulk and surface[13] which may be represented by an equivalent circuit of resistances and capacitances. The observations at any given temperature and frequency is therefore a complex combination of effects from these four elements. The negative values of $\varepsilon'$ observed in the case of x=0.4 above the transition temperature (figure 1) indicates the reduction in the effects from the bulk and domination of surface effects. This is particularly true in measurements with low frequency which is the case in the present measurement. The value of $E_A$ for x = 0.1 and 0.3 samples are in agreement with previous reports on similar half doped $R_{0.5}Ca_{0.5}MnO_3$ (R = Dy, Er and Tb)[12,27] and Bi doped $LaMnO_3$ compounds.[28] The value of activation energy is much smaller in comparison to typical ferroelectrics and dielectric such as $Pb(Zr_{0.2}Ti_{0.8})O_3$ (PZT) where $E_A \approx 0.9eV$.[29] Inset to figures 4 and 5 show the temperature dependence of real part of relative permittivity $\varepsilon'(T)$ at various frequencies for x = 0.1 and 0.3 samples, respectively. The $\varepsilon'(T)$ exhibits a maximum in the range of 180 – 280K, which coincides with the maximum in M(T) (as shown in figure 1). On increasing the frequency, $\varepsilon'$ decreases significantly with the maximum shifting to higher temperature. At 12kHz, maximum value of $\varepsilon'$ for x = 0.1 compound is 320, which is reduced to 225 at 1MHz. Similar to x = 0.1, $\varepsilon'(T)$ for x = 0.3 sample displays a maximum at ~ 280K, with slight dependence on frequency. The maximum value of relative permittivity at 12kHz is 277. It is observed that value of $\varepsilon'$ varies markedly with frequency; however the position of a broad peak described above show a small change with frequency. The $\varepsilon'$ values are a result of both intrinsic and extrinsic effects. The intrinsic contribution is due to the material's bulk response, observed at medium and high frequencies. The extrinsic contribution to $\varepsilon'$ is associated with interfacial polarization produced in grain boundaries and sample-electrode contacts. The extrinsic factors mainly contribute at low frequencies. Equation (1) describes the low temperature behavior but not the presence of maximum at ~ $T_{CO}$. This maximum is observed at higher frequencies as well. Therefore, the presence of maximum in $\varepsilon'(T)$ cannot be attributed to extrinsic factors, and must have intrinsic origin. In figure 6 the imaginary part of relative permittivity $\varepsilon''$ as a function of frequency for x = 0.1 sample at several temperatures is shown. This figure is representative for all the samples having x ≤ 0.4. The $\varepsilon''$ shows a rather high value at low frequencies, which decreases with increasing frequency and decreasing temperature. This behavior is in agreement with earlier reports.[9] The $\varepsilon''$ for a given frequency shows a decrease with increasing temperature, indicating the reduction in $\sigma_{dc}$. In the inset to figure 6 the frequency dependence of imaginary part of relative permittivity $\varepsilon''$ for $La_{0.5}Ca_{0.5-x}Sr_xMnO_3$ (x = 0.1, 0.3 and 0.4) samples at 123K is displayed. As a function of Sr doping at 123K the $\varepsilon''$ shows an increase. This behavior is in agreement with dc resistivity studies reported earlier wherein reduction in resistivity with Sr doping is evident. Figure 7 displays the Argand plane plots for x = 0.1 compound at various temperatures. Similar behavior in Argand plane plot with varying temperature is displayed for x = 0.3 and 0.4 samples. For these samples, single large arc is observed in the temperature range of 123K ≤ T ≤ 293K. This indicates that in this temperature interval and frequency range only intrinsic response is observed. For $La_{0.5}Ca_{0.5}MnO_3$ compound, the intrinsic effects are reported to play an important role in dielectric response.[9] The decrease in size and diameter of the semicircle with increasing temperature reflects the reduction in resistivity of grains.

# CONCLUSION

The dielectric response of polycrystalline La$_{0.5}$Ca$_{0.5-x}$Sr$_x$MnO$_3$ (0.1 ≤ x ≤ 0.4) samples with different magnetic ground state has been studied. It exhibits a maximum coinciding with the maximum observed in M(T) indicating a coupling between the magnetic properties and electric fields. The position of maximum shows slight dependence on frequency. In the antiferromagnetic state the absolute value of $\varepsilon'$ increases with x. The composition (x = 0.4) with A – type antiferromagnetic ordering is found to have a higher $\varepsilon'$ as compared to samples with CE – type ordering despite a higher conductivity in the case of x = 0.4. The frequency dependence of $\varepsilon'$ is described by Debye equation. The obtained values of E$_A$ ~ 0.12eV and do not show significant change with Sr doping up to x ≤ 0.3, while $\tau$ is in the range $2.8 \times 10^{-9} - 6.03 \times 10^{-11}$ s.

**Figure captions**

**Figure 1.** **(a)** Magnetization as a function of temperature at H = 1T and **(b)** the real part of relative permittivity $\varepsilon'$ as a function of temperature at frequency $\nu \approx$ 12kHz for $La_{0.5}Ca_{0.5-x}Sr_xMnO_3$ (x = 0.1, 0.3, and 0.4). The inset to figure 1(b) shows the temperature dependence of imaginary part of relative permittivity $\varepsilon''$ for $La_{0.5}Ca_{0.5-x}Sr_xMnO_3$ (x = 0.1, 0.3, and 0.4) series

**Figure 2.** The Argand plane plot (complex impedance (Z″) versus real part of impedance (Z′)) for impedance behavior for $La_{0.5}Ca_{0.5-x}Sr_xMnO_3$ with x = 0.1, 0.3, and 0.4 at T = 123K.

**Figure 3.** Frequency dependence of real part of relative permittivity $\varepsilon'$ at various temperatures for $La_{0.5}Ca_{0.4}Sr_{0.1}MnO_3$ (x = 0.1) sample. The continuous line is a fit to equation (1).

**Figure 4.** Relaxation time $\tau$ as a function of 1000/T for $La_{0.5}Ca_{0.4}Sr_{0.1}MnO_3$ (x = 0.1) sample. The solid line is the Arrhenius fitting. Inset shows the $\varepsilon'$ versus temperature for x = 0.1 composition at different frequencies.

**Figure 5.** Relaxation time $\tau$ as a function of 1000/T for $La_{0.5}Ca_{0.2}Sr_{0.3}MnO_3$ (x = 0.3) sample. The solid line is the Arrhenius fitting. Inset shows the temperature dependence of $\varepsilon'$ at different frequencies for x = 0.3 sample.

**Figure 6.** Temperature dependence of imaginary part of relative permittivity $\varepsilon''$ for $La_{0.5}Ca_{0.4}Sr_{0.1}MnO_3$ (x = 0.1) sample. This figure is representative for $La_{0.5}Ca_{0.5-x}Sr_xMnO_3$ (x =

0.3 and 0.4) samples. Inset to the figure shows $\varepsilon''$ as a function of frequency at temperature T = 123K for $La_{0.5}Ca_{0.5-x}Sr_xMnO_3$ (x = 0.1, 0.3, and 0.4).

**Figure 7.** The Argand diagram ($Z''$ versus $Z'$) for impedance behavior in $La_{0.5}Ca_{0.4}Sr_{0.1}MnO_3$ (x = 0.1) sample at various temperatures. This figure is representative for $La_{0.5}Ca_{0.5-x}Sr_xMnO_3$ (x = 0.3 and 0.4) samples.

# REFERENCES

1	J. van den Brink and D. Khomskii, J. Phys.: Condens. Matter **20,** 434217 (2008).

2	D. Khomskii, Physics **2,** 20 (2009).

3	M. Fiebig, T. Lottermoser, D. Frohlich, A. V. Goltsev, and R. V. Pisarev, Nature **419,** 418 (2002).

4	T. Kimura, T. Goto, H. Shintani, K. Ishizaka, T. Arima, and Y. Tokura, Nature **426,** 55 (2003).

5	T. Goto, T. Kimura, G. Lawes, A. P. Ramirez, and Y. Tokura, Phys. Rev. Lett. **92,** 257201 (2004).

6	A. M. Kadomtseva, Y. F. Popov, and G. P. Vorob'ev, K. I. Kamilov, A. P. Pyatakov, V. Yu. Ivanov, A. A. Mukhin and A. M. Balbashov, JETP Lett. **81,** 19 (2005).

7	A. M. Kadomtseva, Y. F. Popov, G. P. Vorob'ev, V. Yu. Ivanov, A. A. Mukhin, and A. M. Balbashov, JETP Lett. **82,** 668 (2005).

8	S. Sagar, P. A. Joy, and M. R. Anantharaman, Ferroelectrics **392,** 13 (2009).

9	P. M. Botta, J. Mira, A. Fondado, and J. Rivas, Mater. Lett. **61,** 2990 (2007).

10	G. Giovannetti, S. Kumar, J. van den Brink, and S. Picozzi, Phys Rev Lett **103,** 037601 (2009).

11	R. Seshadri and N. A. Hill, Chem. Mater. **13,** 2892 (2001).

12	Y. Hiramitsu, K. Yoshii, Y. Yoneda, J. Mizuki, A. Nakamura, Y. Shimojo, Y. Ishii, Y. Morii, and N. Ikeda, Jpn. J. Appl. Phys. **46,** 7171 (2007).

13	N. Biskup, A. de Andres, J. L. Martinez, and C. Perca Phys. Rev. B **72,** 024115 (2005).

[14] C. R. Serrao, A. Sundaresan, and C. N. R. Rao, J. Phys.: Condens. Matter **19,** 496217 (2007).

[15] J. Rivas, J. Mira, B. Rivas-Murias, A. Fondado, J. Dec, W. Kleemann, and M. A. Señarís-Rodríguez, Apply Phys Lett **88,** 242906 (2006).

[16] A. M. Lopes, J. P. Araujo, V. S. Amaral, J. G. Correia, Y. Tomioka, and Y. Tokura, Phys. Rev. Lett. **100,** 155702 (2008).

[17] I. Dhiman, A. Das, P. K. Mishra, and L. Panicker, Phys. Rev. B **77,** 094440 (2008).

[18] I. Dhiman, A. Das, R. Mittal, Y. Su, A. Kumar, and A. Radulescu, Phys. Rev. B **81,** 104423 (2010).

[19] P. G. Radaelli, D. E. Cox, M. Marezio, S.-W. Cheong, P. E. Schiffer, and A. P. Ramirez, Phys. Rev. Lett. **75,** 4488 (1995).

[20] S. Mercone, A. Wahl, A. Pautrat, M. Pollet, and C. Simon, Phys. Rev. B **69,** 174433 (2004).

[21] F. Rivadulla, M. A. López-Quintela, L. E. Hueso, C. Jardón, A. Fondado, J. Rivas, M. T. Causa, and R. D. Sánchez, Solid State Commun. **110,** 179 (1999).

[22] M. Sanchez-Andujar, S. Yanez-Vilar, S. Castro-Garcia, and M. A. Senaris-Rodriguez, J. Solid State Chem. **181,** 1354 (2008).

[23] H. Ye, R. B. Jackman, and P. Hing, J. Appl. Phys. **94,** 7878 (2003).

[24] R. C. Kambale, P. A. Shaikh, C. H. Bhosale, K. Y. Rajpure, and Y. D. Kolekar, Smart Mater. Struct. **18,** 085014 (2009).

[25] K. S. Cole and R. H. Cole, J. Chem. Phys. **9,** 341 (1941).

[26] C. C. Wang, G. Z. Liu, M. He, and H. B. Lu, Appl. Phys. Lett. **92,** 052905 (2008).

[27] K. Yoshii, Y. Hiramitsu, Y. Yoneda, Y. Okajima, Y. Nishihata, J. Mizuki, and N. Ikeda, Ferroelectrics **379,** 183 (2009).

[28] Y. J. Wu, Y. Q. Lin, S. P. Gu, and X. M. Chen, Appl Phys A **97,** 191 (2009).

[29] C. S. Ganpule, A. L. Roytburd, V. Nagarajan, B. K. Hill, S. B. Ogale, E. D. Williams, R. Ramesh, and J. F. Scott, Phys. Rev. B **65,** 014101 (2001).

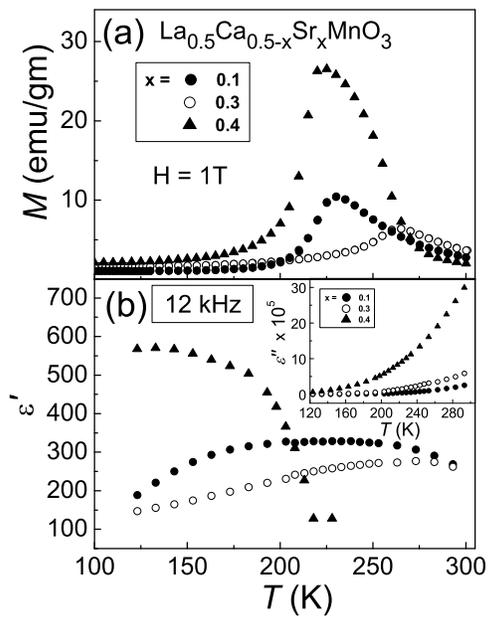

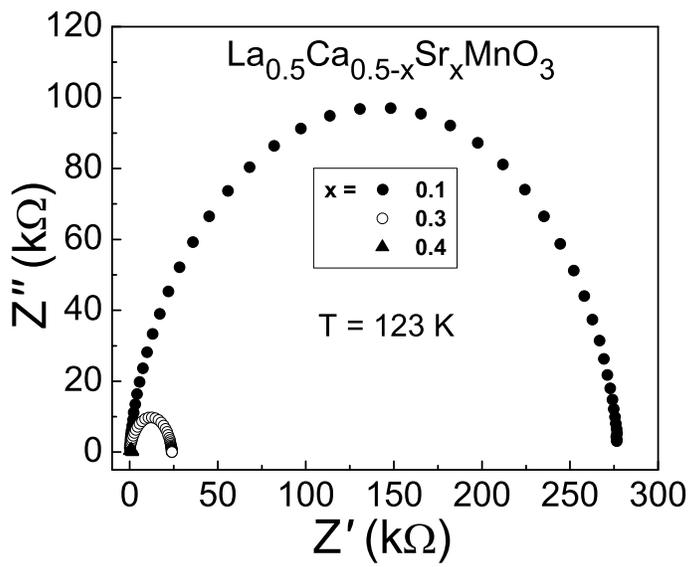

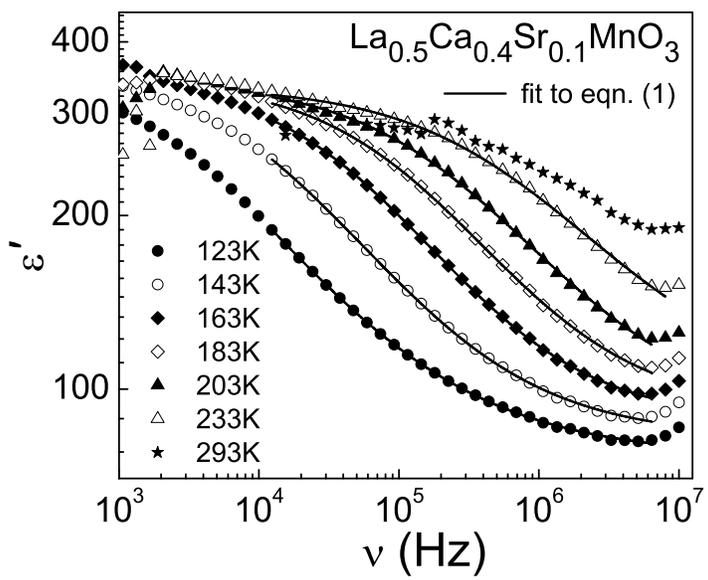

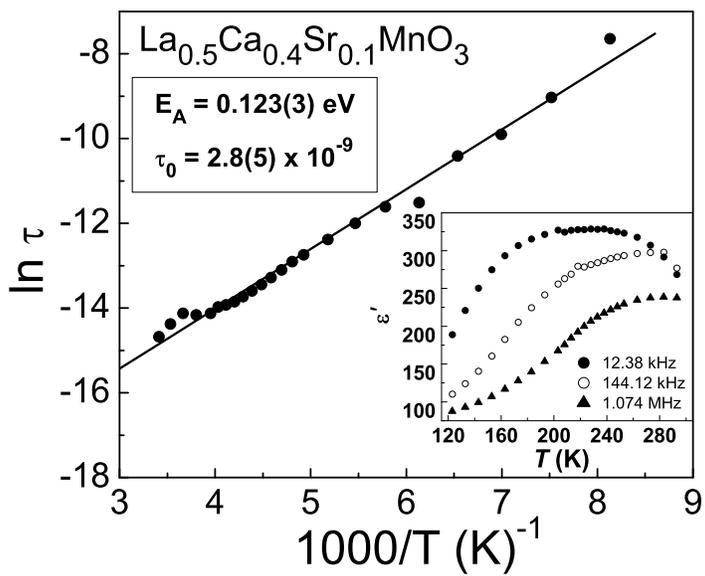

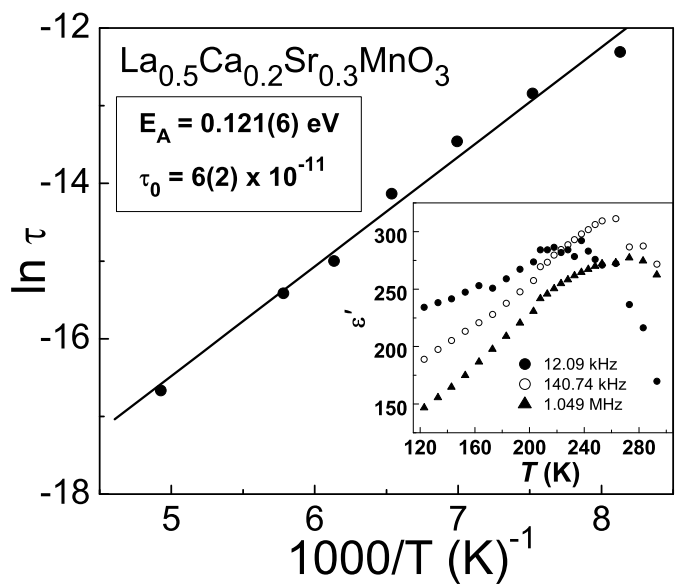

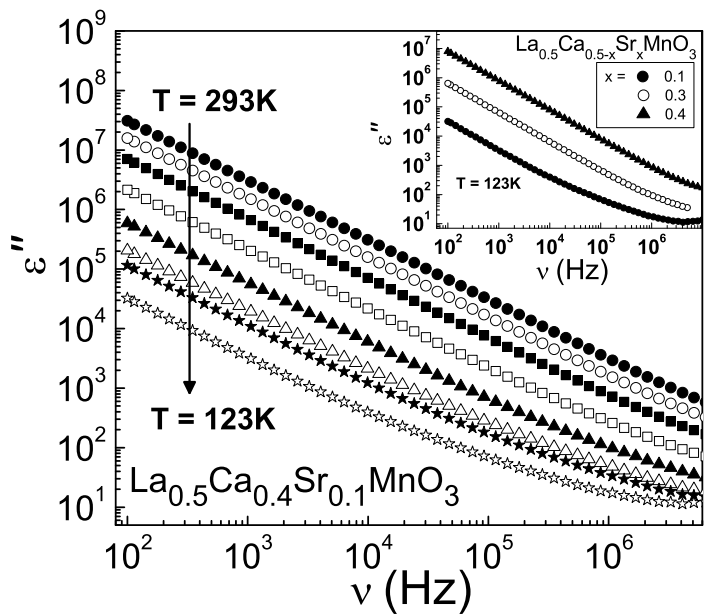

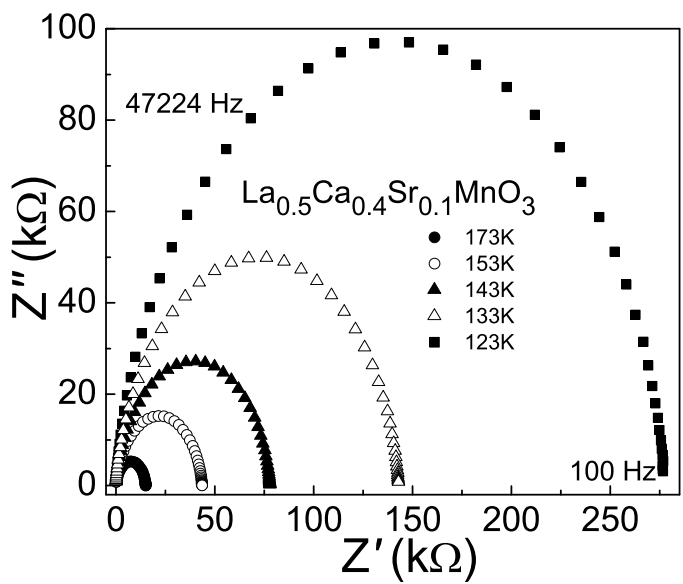